\documentclass[12pt]{article}
\pdfoutput=1

\usepackage{draft} 
\usepackage{hyperref}
\usepackage{graphicx,color,subfig}
\usepackage{cite}
\usepackage{mciteplus}
\usepackage{skak}
\usepackage{empheq}
\usepackage{anyfontsize}
\usepackage{amsthm}

\usepackage{comment}
\usepackage{algorithm}
\usepackage{algpseudocode}

\DeclareFontFamily{OT1}{pzc}{}
\DeclareFontShape{OT1}{pzc}{m}{it}{<-> s * [1.10] pzcmi7t}{}
\DeclareMathAlphabet{\mathpzc}{OT1}{pzc}{m}{it}

\def\be#1\ee{\begin{align}#1\end{align}}

\begin{document}

\unitlength = .8mm

\begin{titlepage}

\begin{center}

\hfill \\
\hfill \\
\vskip 1cm

\title{Nonequilibrium Phase Transitions in\\Large $N$ Matrix Quantum Mechanics}

\author{Minjae Cho}

\address{
Leinweber Institute for Theoretical Physics, University of Chicago, Chicago, IL 60637, USA
}

\email{cho7@uchicago.edu}

\end{center}

\abstract{It is believed that the theory of quantum gravity describing our universe is unitary. Nonetheless, if we only have access to a subsystem, its dynamics is described by nonequilibrium physics. Motivated by this, we investigate the planar limit of large $N$ ungauged one-matrix quantum mechanics obeying the Lindblad master equation with dissipative jump terms, focusing on the existence, uniqueness, and properties of steady states that signal nonequilibrium phase transitions. In the first class of examples, where potentials are unbounded from below, we study nonequilibrium critical points above which strong dissipation allows for the existence of normalizable steady states that would otherwise not exist. In the second class of examples, termed matrix quantum optics, we find evidence of nonequilibrium phase transitions analogous to those recently reported in the quantum optics literature for driven-dissipative Kerr resonators. Preliminary results on two-matrix quantum mechanics are also presented. We implement bootstrap methods to obtain concrete and rigorous results for the nonequilibrium steady states of matrix quantum mechanics in the planar limit.}

\end{titlepage}

\eject

\begingroup
\hypersetup{linkcolor=black}

\tableofcontents

\endgroup

\section{Introduction}
Nonequilibrium physics appears ubiquitously in nature. One outstanding class of examples is provided by open systems, where coupling to the environment introduces dissipation. In the absence of an energy functional, there is no notion of equilibrium. Instead, nonequilibrium physics presents its own list of intriguing questions.

One of the fundamental questions concerns the existence, uniqueness, and properties of steady states, which remain invariant under time evolution, as they govern the late-time behavior of the system. In particular, the properties of steady states can change dramatically at nonequilibrium critical points. There are numerous examples of nonequilibrium phase transitions in both classical and quantum many-body systems, ranging from classical directed percolation\footnote{See \cite{Hinrichsen01112000} and references therein for an introduction to nonequilibrium phase transitions in the context of classical stochastic processes.} to quantum transitions of an information-theoretic nature \cite{PhysRevB.98.205136,PhysRevX.9.031009,PRXQuantum.4.030317}. Notably, many intriguing examples in quantum systems exhibit transitions between two distinct phases: one where quantum coherence is maintained, and another where quantum dissipation dominates.

Regarding quantum gravity as a quantum system, we can consider the case where we have access only to a subsystem, which then becomes an open quantum system. A fascinating example is a black hole that evaporates due to its interactions with the environment, with special cases having been studied in recent years \cite{Penington:2019npb,Almheiri:2019psf}. Another important example is the open effective theory for long-wavelength modes in cosmology \cite{Salcedo:2024smn,Li:2025azq,Salcedo:2025ezu}. An interesting question is whether nonequilibrium phase transitions can arise in open quantum gravity theories. Such transitions are inherently different in nature from the more familiar equilibrium phase transitions in quantum gravity \cite{cmp/1103922135,ATICK1988291,Horowitz:1996nw,Horowitz:1997jc,Aharony:2003sx}.

In this work, we study the nonequilibrium physics of large $N$ matrix quantum mechanics (MQM).\footnote{To the best of our knowledge, such an open MQM system was first studied in \cite{Cao:2022fof}. For an introduction to ordinary MQM at equilibrium, we refer readers to \cite{Klebanov:1991qa,Marino_2015}.} A general MQM system is expected to be dual to a quantum gravity theory. Well-known examples include $c=1$ MQM (see \cite{Klebanov:1991qa} for a review), type 0B MQM \cite{Takayanagi:2003sm,Douglas:2003up}, and Banks-Fischler-Shenker-Susskind (BFSS) MQM \cite{Banks:1996vh}. Given their known string and M-theory dual descriptions, there is no strong motivation to consider such explicit examples as subsystems of larger quantum systems. We may choose to couple them to the environment regardless, or instead consider their subsystems, which still take the form of MQM but are coupled to the environment. In either case, we can ask questions about the existence and properties of steady states of the subsystem.

Specifically, we focus on MQM subject to the Gorini–Kossakowski–Sudarshan–Lindblad master equation \cite{1976CMaPh..48..119L,1976JMP....17..821G}, or Lindblad equation for short, which describes the most general Markovian, completely positive, and trace-preserving time evolution of density matrices $\rho$:\footnote{For a brief introduction to the Lindblad equation, see \cite{10.1063/1.5115323}. We also remark that the Lindblad dynamics of another popular large $N$ quantum system, the Sachdev–Ye–Kitaev (SYK) model \cite{PhysRevLett.70.3339,Kitaev}, has been studied in \cite{PhysRevResearch.4.L022068,PhysRevB.106.075138}.}
\ie\label{eqn:LindbladMaster}
{d\over dt}\rho={\cal L}(\rho)=-i[H,\rho]+\sum_K\gamma_K\left(L_K\rho L_K^\dagger-{1\over2}\{L^\dagger_K L_K,\rho\} \right).
\fe
Here, $\rho$ and $H$ denote the density matrix and Hamiltonian of the subsystem, respectively, while $L_K$ (for appropriate indices $K$) are linear operators on the subsystem Hilbert space, referred to as jump operators, which represent dissipation with damping rates $\gamma_K \geq 0$. The time evolution operator ${\cal L}$ is called the Liouvillian superoperator.

Of course, whether the Lindblad equation provides a good effective description for an open quantum system depends on the specifics of the setup of interest. In particular, the Markovian property may not hold, and the environment may build a nontrivial memory of the subsystem's evolution, in which case the Lindblad equation is not applicable. Finding or engineering specific setups in quantum gravity where the Lindblad equation provides a good effective description is beyond the scope of the current work. In this work, we will assume that the Lindblad equation provides an appropriate description of the subsystem of interest. Our primary interest here lies in exploring possible nonequilibrium phase transitions in quantum gravity, and the Lindblad equation offers an amenable setup for such investigations in the context of MQM.

Our primary example is one-matrix quantum mechanics (1-MQM) with $N\times N$ Hermitian matrices $X$ and $P$ subject to the canonical commutation relations
\ie
[X_{ij},P_{kl}]=i\delta_{il}\delta_{jk},~~~i,j,k,l=1,2,\cdots,N,
\fe
where the Hamiltonian $H$ is a single-trace operator. For example, we may consider Hamiltonians of the form\footnote{Our convention for matrix trace normalization is $\Tr X=\sum_{i=1}^N X_{ii}$.}
\ie
H=\Tr\left({1\over2}P^2+N~V\left({X\over\sqrt{N}}\right)\right),
\fe
where the $N$-scaling of the potential term $V$ is chosen so that the Hamiltonian expectation value scales as $\langle H \rangle \sim N^2$ in the 't Hooft large $N$ limit. To ensure that the dissipative effects are comparable in magnitude to the Hamiltonian contribution, the natural scaling for the jump terms is $\sum_K\gamma_K\langle L_K^\dagger L_K \rangle \sim N^2$, which we assume throughout this work.

The Hamiltonian has a $U(N)$ symmetry under which $X$ and $P$ transform in the adjoint representation. In the absence of dissipation, there are two versions of the theory one can consider: (1) gauged MQM, where the Hilbert space is restricted to $U(N)$ singlets, and (2) ungauged MQM, where there is no such restriction.

We begin by considering the planar gauged MQM. In the gauged MQM, well-defined linear operators are $U(N)$ singlets of the trace form $\Tr(\cdots)$, in which case the jump operators $L_K$ are also products of traces. Consider the time evolution of operator expectation values $\langle \cal O \rangle = \text{tr}\left(\rho {\cal O} \right)$, where $\text{tr}$ denotes the trace over the Hilbert space, as governed by (\ref{eqn:LindbladMaster}):
\ie\label{eqn:LindbladOperatorEvolution}
{d\over dt}\langle {\cal O}\rangle &= i\langle[H,{\cal O}]\rangle+\sum_K\gamma_K\left(\langle L_K^\dagger{\cal O}L_K\rangle-{1\over2}\langle L_K^\dagger L_K{\cal O}\rangle-{1\over2}\langle {\cal O}L_K^\dagger L_K\rangle\right)
\\
&=i\langle[H,{\cal O}]\rangle+{1\over2}\sum_K\gamma_K\left(\langle[L^\dagger_K,{\cal O}]\rangle\langle L_K\rangle+\langle [{\cal O},L_K]\rangle\langle L^\dagger_K\rangle\right),
\fe
where we used large $N$ factorization in the second line. We observe that, in general, there is nontrivial Lindblad dissipation for the planar gauged MQM, where the time evolution is nonlinear in the operator expectation values.\footnote{We thank the anonymous referee of JHEP for pointing out that Lindblad dissipation may be nontrivial in the planar gauged MQM.} For example, if $\mathcal{O}$ and $L_K$ are single-trace operators, ${d\over dt}\langle \mathcal{O} \rangle$ contains quadratic terms in the expectation values, even before taking into account any trace-cyclicity considerations to be discussed later (e.g. (\ref{eqn:cyclicity})). For this reason, analyzing the Lindblad dissipation of the planar gauged MQM is a highly involved task, which we will explore in a separate work \cite{unpubl1}.

In this work, we instead consider ungauged MQM, where $L_K$ are in non-singlet representations of $U(N)$. As a concrete example, define the adjoint annihilation and creation operators
\ie\label{eqn:adjointCreationAnnihilation}
A_{ij}={1\over\sqrt{2}}(X_{ij}+i P_{ij}),~~~B_{ij}=A_{ji}^\dagger={1\over\sqrt{2}}(X_{ij}-i P_{ij}),
\fe
obeying
\ie
[A_{ij},B_{kl}]=\delta_{il}\delta_{jk},
\fe
and take $L_K$ to be $A_{ij}$, where $\gamma_K$ is independent of $K$, leading to
\ie\label{eqn:LindbladSingleAdjEmissionDensity}
{d\over dt}\rho=-i[H,\rho]+\gamma\left(-{1\over2}\{\Tr(BA),\rho\}+\sum_{i,j}A_{ij}\rho B_{ji} \right),
\fe
or equivalently,
\ie\label{eqn:LindbladSingleAdjEmission}
{d\over dt}\langle {\cal O}\rangle = i\langle[H,{\cal O}]\rangle+\gamma\left({1\over2}\langle[\Tr(BA),{\cal O}]\rangle+\sum_{i,j}\langle B_{ij}[{\cal O},A_{ji}]\rangle\right).
\fe
Now, the dissipation term is not manifestly nonlinear in the expectation values and is therefore more amenable to practical investigation compared to (\ref{eqn:LindbladOperatorEvolution}).\footnote{Although (\ref{eqn:LindbladSingleAdjEmission}) will involve nonlinear terms in the expectation values at the planar limit once trace cyclicity and operator algebra are taken into account (see section \ref{sec:bootstrap}), its degree of non-convexity is nevertheless much milder than that of (\ref{eqn:LindbladOperatorEvolution}).} This dissipation can be regarded as a spontaneous emission of a single MQM quantum to a heat sink. This quantum is in the adjoint representation, sometimes referred to as the (off-shell) long strings \cite{1980JMP....21.1103M,Maldacena:2005hi,Maldacena:2018vsr,Balthazar:2018qdv}. (\ref{eqn:LindbladSingleAdjEmissionDensity}) respects the $U(N)$ symmetry in that the time evolution and the $U(N)$ transformations commute with each other:
\ie
{\cal L}([G_{ij},\rho])=[G_{ij},{\cal L}(\rho)],
\fe
where $G_{ij}=i[X,P]_{ij}+N\delta_{ij}$ is the $U(N)$ generator.

Steady states $\rho$ are defined by ${d\over dt}\rho=0$, or equivalently via their expectation values as
\ie\label{eqn:steadyStateCondition}
{d\over dt}\langle {\cal O}\rangle = i\langle[H,{\cal O}]\rangle+\gamma\left({1\over2}\langle[\Tr(BA),{\cal O}]\rangle+\sum_{i,j}\langle B_{ij}[{\cal O},A_{ji}]\rangle\right)=0,
\fe
for all operators ${\cal O}$. If steady states exist, then there must be at least one that transforms as a $U(N)$ singlet, i.e., $[G_{ij},\rho]=0,~\forall i,j$. This does not necessarily imply that such a $\rho$ is an ensemble composed only of $U(N)$ singlet states in the Hilbert space, which would further require $G_{ij}\rho=0$—a condition that fails in many examples we will consider. In the language of \cite{deGroot2022symmetryprotected}, density matrices satisfying $G_{ij}\rho=0$ are said to be strongly symmetric, while those satisfying $[G_{ij},\rho]=0$ but not $G_{ij}\rho=0$ are weakly symmetric under $U(N)$. Suppose a steady state $\rho$ satisfies $[G_{ij}, \rho] = 0$. If $\rho$ is pure, then it is strongly symmetric. Equivalently, if it is weakly but not strongly symmetric, then it must be a mixed state. Therefore, if there are only weakly symmetric steady states, we conclude that all steady states must be mixed states, a statement that has been useful in studying nonequilibrium phase transitions in recent literature (see e.g. \cite{deGroot2022symmetryprotected, PRXQuantum.4.030317}).

We explore two central questions. Given an MQM Hamiltonian and a value $\gamma$ for the strength of dissipation, does a normalizable steady state exist in the planar limit? If it does, is it unique, and what are its properties? We will find examples where the answers to these questions depend on the strength of the dissipation $\gamma$. For examples with Hamiltonians that have unbounded potentials, there is a nonequilibrium critical point $\gamma_c$ such that a normalizable steady state exists if and only if $\gamma \geq \gamma_c$. This suggests the existence of stationary quantum gravity backgrounds in the presence of a heat sink, which otherwise cannot exist.

In other examples, called matrix quantum optics, the number of steady states changes from one to multiple across nonequilibrium critical points, and the order parameter exhibits nonanalytic behavior. Matrix quantum optics describes a cavity containing adjoint quanta subject to two competing effects: spontaneous emission of adjoint quanta to the environment, and pumping of adjoint quanta from external sources. Two effects together may lead to interesting nonequilibrium phase transitions. These examples are motivated by recently discovered nonequilibrium phase transitions in quantum optics for driven-dissipative Kerr resonators \cite{PhysRevA.94.033841,PhysRevA.95.012128,PhysRevA.98.042118,Beaulieu2025}, whose thermodynamic limit is reminiscent of the large $N$ limit of MQM systems.

Except for special cases, computing the properties of the steady states obeying (\ref{eqn:steadyStateCondition}) is not straightforward, even for 1-MQM, since the model is ungauged and thus the picture of non-interacting fermions applicable to gauged 1-MQM does not apply. There is also no obvious perturbative scheme, and Monte Carlo simulations based on a real-valued energy functional are simply not available. At finite $N$, one may truncate the Hilbert space to obtain estimates. However, the computational cost grows rapidly as $N$ and the truncation level increase.

For this reason, we use the bootstrap method based on the positivity of $\rho$:
\ie
\langle {\cal O}^\dagger{\cal O}\rangle\geq0,
\fe
for any operator ${\cal O}$. When combined with the steady state condition (\ref{eqn:steadyStateCondition}), we can obtain rigorous results for the steady states in the strict planar limit. Recent studies of bootstrap methods for equilibrium MQM in the planar limit have proved to be very successful, yielding both rigorous and precise bounds, or demonstrating the absence of density matrices obeying certain constraints \cite{Lin:2020mme,Han:2020bkb,Lin:2023owt,Cho:2024kxn,Lin:2024vvg,Lin:2025srf}. Furthermore, bootstrap methods for nonequilibrium physics have very recently been established in both classical stochastic processes \cite{Cho:2025dgc} and quantum Lindblad dynamics \cite{Robichon:2024apx,Mortimer:2024fuu}, and have successfully addressed nonequilibrium phase transitions.

In this work, the bootstrap will allow us to extract highly nontrivial information about the nature of nonequilibrium physics in MQM Lindblad systems. Even though bootstrap provides a rigorous proof of non-existence, rather than existence, of certain steady states, we will find convincing evidence of nonequilibrium phase transitions from nearly nonanalytic behaviors of bootstrap bounds.

This paper is organized as follows. We begin with simple, exactly solvable examples in section \ref{sec:warmup}, where we observe in particular that normalizable steady states may exist for unbounded potentials if the dissipation is strong enough. In section \ref{sec:unbounded}, we provide a brief review of the bootstrap method and implement it to study nonequilibrium critical points of 1-MQM with an inverted double-well potential. The bootstrap produces lower bounds on the minimal value of the dissipation strength above which normalizable steady states exist. A similar, albeit preliminary, analysis is carried out for two-matrix quantum mechanics (2-MQM) with an unbounded potential. We then proceed to the matrix quantum optics examples in section \ref{sec:MQO}, where a matrix generalization of the driven-dissipative Kerr resonator in quantum optics is introduced. Bootstrap bounds on matrix quantum optics exhibit nonanalytic behavior, suggesting the existence of nonequilibrium phase transitions. We conclude with further discussion and future prospects in section \ref{sec:discussion}. Appendix \ref{apx:kerrBoot} discusses the bootstrap approach to the driven-dissipative Kerr resonator in quantum optics, demonstrating its effectiveness in diagnosing nonequilibrium phase transitions.

\textit{Note added:} This paper is submitted in coordination with \cite{shinsei}, which explores related aspects of the stabilization of an unbounded potential via Lindblad dissipation in the bosonic SYK model.

\section{Warmup: exactly solvable cases}\label{sec:warmup}
In this section, we discuss simple examples of 1-MQM Lindblad equations (\ref{eqn:LindbladSingleAdjEmissionDensity}) with exactly solvable steady states. These examples are solvable for any value of $N$.

\subsection{Fock vacuum case}
We start by defining the number operator ${\cal N}=\Tr(BA)$, where $A$ and $B$ are the adjoint annihilation and creation operators introduced in (\ref{eqn:adjointCreationAnnihilation}). Assume that the Hamiltonian $H$ commutes with ${\cal N}$. An example is
\ie\label{eqn:trivialEx1Ham}
H=\Tr\left(BA+{\omega\over N} B^2A^2 \right).
\fe
For such Hamiltonians, the Lindblad equation (\ref{eqn:LindbladSingleAdjEmission}) leads to
\ie\label{eqn:ndissipation}
{d\over dt}\langle{\cal N}\rangle = -\gamma\langle{\cal N}\rangle.
\fe
Therefore, the steady state must satisfy $\langle{\cal N}\rangle=0$, which pins it down to $\rho=|0\rangle\langle0|$, where $|0\rangle$ is the Fock vacuum defined by $A_{ij}|0\rangle=0$ for all $i$ and $j$. It is a pure state and therefore strongly symmetric under $U(N)$.

Later in section \ref{sec:MQO}, we will add a one-quantum pumping term $\sqrt{N}\Tr \left(A+B\right)$ or a two-quantum pumping term $\Tr(A^2+B^2)$ to the Hamiltonian (\ref{eqn:trivialEx1Ham}), in which case the steady state is no longer given by the Fock vacuum and is not exactly solvable.

\subsection{Quadratic case}\label{sec:quadraticEx}
When the Lindblad equation (\ref{eqn:LindbladMaster}) is quadratic—in the sense that the Hamiltonian is quadratic and the jump operators are linear—it is exactly solvable. In fact, the ungauged 1-MQM case (\ref{eqn:LindbladSingleAdjEmissionDensity}) with such a quadratic structure corresponds to $N^2$ decoupled copies of single bosonic particle quantum mechanical systems, and it therefore suffices to study a single copy—a case that has been extensively studied in the quantum optics literature.

Consider a single bosonic particle, whose position, momentum, annihilation, and creation operators we denote by $x$, $p$, $a={x+ip\over\sqrt{2}}$, and $a^\dagger={x-ip\over\sqrt{2}}$, respectively. We consider the following Lindblad equation studied in \cite{Downing2023}:
\ie
{d\over dt}\rho=-i[h,\rho]+\gamma\left(a\rho a^\dagger-{1\over2}\{a^\dagger a,\rho\} \right),
\fe
with the Hamiltonian
\ie
h={1\over2}\left(p^2-x^2\right)=-{1\over2}\left((a^\dagger)^2+a^2\right).
\fe
Since the potential is unbounded from below, there is no normalizable steady state when $\gamma=0$. In contrast, nonzero $\gamma$ introduces a heat sink, which may suppress the unbounded nature of the potential and thus allow for a normalizable steady state. In fact, the Hamiltonian can be interpreted as a two-quanta driving term, placing the current problem within a well-studied class of parametrically driven oscillators subject to quantum noise.\footnote{See, e.g., \cite{RevModPhys.82.1155} for a review on quantum amplification and noise. We thank Aashish Clerk and Andrew Pocklington for introducing us to this topic.}

Since the system is quadratic, we can solve for the steady state exactly. Using the standard $P$-representation \cite{Drummond_1980}, the unique solution $\rho$ to the steady state equation ${d\over dt}\rho=0$, with a normalization such that $\rho$ has unit trace when it is well-defined, is given by
\ie\label{eqn:singleBosonSteady}
\rho={\sqrt{\gamma^2-4}\over2\pi}\int_{-\infty}^\infty dt \int_{-\infty}^\infty ds ~e^{-{\gamma\over2}(t^2+s^2)+2st}~{|\omega t\rangle_c~{}_c\langle\omega s| \over {}_c\langle\omega s|\omega t\rangle_c},
\fe
where $\omega=e^{i\pi \over4}$ and $|\alpha\rangle_c$ is the normalized coherent state satisfying $a|\alpha\rangle_c=\alpha|\alpha\rangle_c$. However, $\rho$ given by (\ref{eqn:singleBosonSteady}) is a positive semidefinite density matrix of finite norm if and only if $\gamma > 2$. One quick consistency check is given by the expectation value of the number operator $n = a^\dagger a$:
\ie
\langle n\rangle = {\sqrt{\gamma^2-4}\over2\pi}\int_{-\infty}^\infty dt \int_{-\infty}^\infty ds ~st~e^{-{\gamma\over2}(t^2+s^2)+2st}={2\over\gamma^2-4},
\fe
which is nonnegative and finite only if $\gamma > 2$.

This simple example therefore exhibits a nonequilibrium critical point at $\gamma = \gamma_c = 2$.\footnote{In \cite{Downing2023}, such critical points were referred to as exceptional points.
} For $\gamma \leq \gamma_c$, there does not exist a normalizable steady state. In contrast, for $\gamma > \gamma_c$, there is a unique normalizable steady state, which is a mixed state, as can be seen from its purity ${\sqrt{\gamma^2-4}\over\gamma} < 1$. Its energy expectation value is given by $\langle h \rangle = 0$, independent of the value of $\gamma$.

The 1-MQM version is given by (\ref{eqn:LindbladSingleAdjEmissionDensity}) with the Hamiltonian
\ie\label{eqn:c1Hamiltonian}
H={1\over2}\Tr(P^2-X^2)=-{1\over2}\Tr(B^2+A^2).
\fe
The unique steady state is simply the $N^2$-fold tensor product of $\rho$ in (\ref{eqn:singleBosonSteady}), whose expectation values of the number operator ${\cal N}$ and the Hamiltonian are given by
\ie
{\langle{\cal N}\rangle\over N^2}={2\over\gamma^2-4},~~~{\langle H\rangle\over N^2} = 0.
\fe
Again, there is a nonequilibrium critical point at $\gamma = \gamma_c = 2$, above which a normalizable steady state exists.

The Hamiltonian (\ref{eqn:c1Hamiltonian}) is nothing but that of $c=1$ and type 0B MQM. However, the vacua over which these theories are defined are quite different from the normalizable states of interest in this work. For the gauged model, the vacua of $c=1$ and type 0B MQM are obtained in the double-scaling limit—rather than merely the ’t Hooft limit—by filling the Fermi levels on one side and on both sides of the top of the potential, respectively. The observables of interest are the scattering amplitudes of fluctuations around the Fermi surface~\cite{Moore:1991zv}, which match the dual string theory amplitudes~\cite{DiFrancesco:1991ocm,DiFrancesco:1991daf,Balthazar:2017mxh}.

The adjoint sector of the ungauged model also allows for scattering states, called long strings, which match the open strings on FZZT branes in certain limits \cite{Maldacena:2005hi,Balthazar:2018qdv}. In such discussions, long strings are regarded as fluctuations around the aforementioned singlet vacua. The investigation of the precise dual string theory description of the steady state at $\gamma > \gamma_c$ is beyond the scope of this work. Given that both $c=1$ and type 0B string theories in their current formulations describe perturbative scattering string states dual to fluctuations of the Fermi surface in the double-scaling limit, it remains unclear whether there is a practical tool to describe the string background dual to the steady state. Firstly, there does not appear to be a natural double-scaling limit for the normalizable steady state, since there is no infinite Fermi sea to fill. Secondly, the normalizable steady state is obtained in the ’t Hooft limit, whereas perturbative string theory computations produce quantities only in the double-scaling limit. On another note, the non-singlet sector of MQM has been argued to be related to black holes in these string theories \cite{Maldacena:2005hi,Betzios:2017yms,Betzios:2022pji,Ahmadain:2022gfw}. It is therefore very interesting to study the Lindblad dynamics of MQM in the double-scaling limit, as opposed to the ’t Hooft limit considered in this work. We leave a detailed investigation of this question for future work.

\section{Unbounded potential examples}\label{sec:unbounded}
We now consider unsolvable MQM Lindblad equations. Before turning to specific examples, we first introduce the bootstrap method, which will be used to obtain rigorous and precise bounds on the expectation values of steady states. Such a method for bootstrapping steady states of Lindblad equations was recently introduced in \cite{Robichon:2024apx,Mortimer:2024fuu}, and we apply it to MQM Lindblad equations.

\subsection{Review of the bootstrap method}\label{sec:bootstrap}
The bootstrap method for equilibrium MQM was first implemented in \cite{Polchinski:1999br} and revisited more systematically in \cite{Han:2020bkb}. For ungauged MQM, it was recently implemented in \cite{Cho:2024kxn}, which we closely follow in this work.

MQM Hamiltonians are assumed to be quartic in the adjoint variables $X$ and $P$. Consider ``open words'' of $X$ and $P$, which are adjoint operators constructed by taking products of $X$ and $P$. An example of an open word is $X^2P^2XPX^3P$. Define ${\cal B}_L$ to be the set of open words whose lengths are at most $L$. For example, ${\cal B}_2 = \{1, X, P, X^2, XP, PX, P^2\}$.

Single-trace expectation values $\langle\Tr{\cal O}\rangle$ for ${\cal O} \in {\cal B}_L$ are the bootstrap variables. Since we are interested in the planar limit, large $N$ factorization, 
\ie
\langle\Tr{\cal O}_1\Tr{\cal O}_2\cdots\Tr{\cal O}_n\rangle = \langle\Tr{\cal O}_1\rangle\langle\Tr{\cal O}_2\rangle\cdots\langle\Tr{\cal O}_n\rangle,
\fe
for open words ${\cal O}_a$, implies that multi-trace expectation values are not independent from the single-trace expectation values.

The first bootstrap constraint is the positivity condition $\langle \Tr({\cal O}^\dagger{\cal O})\rangle \geq 0$ for all adjoint operators ${\cal O}$. It should hold for any valid density matrix. To truncate to a finite basis, we define the matrix
\ie
{\cal M}^{(L)}_{a,b} = \langle \Tr({\cal O}_a^\dagger{\cal O}_b)\rangle, \qquad {\cal O}_a, {\cal O}_b \in {\cal B}_{L/2}.
\fe
When $L$ is odd, ${\cal B}_{L/2}$ is understood to be the same as ${\cal B}_{(L-1)/2}$. Positivity of the density matrix then implies that ${\cal M}^{(L)}$ is a positive semidefinite matrix:
\ie\label{eqn:bootstrapPositivity}
{\cal M}^{(L)} \succeq 0.
\fe

The second ingredient is the cyclicity of traces. Since $X$ and $P$ do not commute, the trace of an open word is cyclic only up to commutator terms. For example,
\ie\label{eqn:cyclicity}
\langle \Tr(P^2X^3P)\rangle = \langle \Tr(P^3X^3)\rangle + iN\langle \Tr(P^2X^2)\rangle + i\langle \Tr(X)\rangle\langle \Tr(P^2X)\rangle + i\langle \Tr(P^2)\rangle\langle \Tr(X^2)\rangle.
\fe
In deriving such cyclicity relations, we organize terms into products of traces so that there are no open indices. Note that cyclicity relations are generally nonlinear in the single-trace expectation values.

The third ingredient is the reality constraint $\langle {\cal O}^\dagger \rangle = \langle {\cal O} \rangle^*$, which holds for any operator ${\cal O}$. We treat these as linear constraints on the bootstrap variables.

Bootstrap constraints introduced so far apply to expectation values of any density matrix in ungauged 1-MQM. Since we are interested in the steady states of the Lindblad master equation, additional constraints corresponding to steady states should be imposed. We call these the steady state constraints. The first set of such constraints is given by (\ref{eqn:steadyStateCondition}) for single-trace operators ${\cal O}$, i.e., ${d\over dt}\langle\Tr{\cal O'}\rangle = 0$ for any open word ${\cal O'}$. The second set of constraints is given by (\ref{eqn:steadyStateCondition}) with ${\cal O}$ being a multi-trace operator. A simplification in the planar limit is that large $N$ factorization implies
\ie\label{eqn:multiTraceSteadyState}
{}&\bigg\langle\Tr{\cal O}_1\cdots\Tr{\cal O}_{n-1}\left(i[H,\Tr{\cal O}_n]+\gamma\left({1\over2}[\Tr(BA),\Tr{\cal O}_{n}]+\sum_{i,j}B_{ij}[\Tr{\cal O}_n,A_{ji}]\right)\right)\bigg\rangle
\\
&=\langle\Tr{\cal O}_1\rangle\cdots\langle\Tr{\cal O}_{n-1}\rangle\bigg\langle\left(i[H,\Tr{\cal O}_n]+\gamma\left({1\over2}[\Tr(BA),\Tr{\cal O}_{n}]+\sum_{i,j}B_{ij}[\Tr{\cal O}_n,A_{ji}]\right)\right)\bigg\rangle = 0,
\fe
for steady states, from which ${d\over dt}\langle\Tr{\cal O}_1\cdots\Tr{\cal O}_n\rangle = 0$ follows straightforwardly.

Among the cyclicity relations and steady state constraints, we will restrict ourselves to those that depend only on single- and double-trace expectation values. The bootstrap problem is then quadratic in the single-trace expectation values. Such a quadratic problem allows for a convex relaxation, taking the form of a semidefinite programming (SDP) problem, as we now explain.

Suppose we restrict to constraints that involve words only up to a maximal length $L$. Using the bootstrap constraints mentioned above, except for the positivity constraint, we first solve for as many double-trace expectation values as possible in terms of single-trace expectation values. For the remaining double-trace expectation values, say $\langle\Tr{\cal D}_u\rangle\langle\Tr{\cal D}_v\rangle$ for appropriate indices $u, v$ and Hermitian single-trace operators $\Tr{\cal D}_u$ and $\Tr{\cal D}_v$, we implement the convex relaxation introduced in \cite{Kazakov:2021lel,Kazakov:2022xuh,Lin:2025srf} as follows. Introduce new bootstrap variables $w_{u,v}$ to replace the double-trace expressions $\langle\Tr{\cal D}_u\rangle\langle\Tr{\cal D}_v\rangle$. Instead of imposing $w_{u,v} = \langle\Tr{\cal D}_u\rangle\langle\Tr{\cal D}_v\rangle$, we treat $w_{u,v}$ as a matrix element of a matrix $w$, introduce a column vector $y_u = \langle\Tr{\cal D}_u\rangle$, and impose that the following matrix ${\cal W}^{(L)}$ is positive semidefinite:
\ie\label{eqn:convexRelaxation}
{\cal W}^{(L)}=
\begin{pmatrix}
1 & y^T \\
y & w
\end{pmatrix} \succeq 0.
\fe

All bootstrap constraints are convex after the relaxation. We consider objective functions of the form $\langle \Tr{\cal Q} \rangle$, where $\Tr{\cal Q}$ is Hermitian and ${\cal Q}$ is a sum of words of lengths smaller than $L$. The final form of the bootstrap problem for steady states involving words of lengths up to $L$ is given by the following SDP problem:\footnote{SDP is a convex optimization problem of finding the minimum (or maximum) of $\sum_{i=1}^m c_i x_i$ subject to the matrix positivity constraint $\textbf{X}=\sum_{i=1}^m F_i x_i - F_0 \succeq 0$, where $x_i$ are real variables, $c_i$ are real constants, and $F_0$ and $F_i$ are real-symmetric constant matrices. For an introduction to SDP, we refer the reader to \cite{doi:10.1137/1038003}.}
\ie\label{SDPDefinition}
{}&\textbf{SDP($L$): } \text{ minimize (maximize) } \langle\Tr{\cal Q}\rangle \text{ subject to} \\
&\text{1. Positivity: } {\cal M}^{(L)} \succeq 0 \\
&\text{2. Normalization: } \langle 1 \rangle = 1 \\
&\text{3. Cyclicity of traces for } \langle\Tr {\cal O}\rangle,\ {\cal O} \in {\cal B}_L \\
&\text{4. Reality constraints for } \langle\Tr {\cal O}\rangle,\ {\cal O} \in {\cal B}_L \\
&\text{5. Single-trace steady state constraints: } {d\over dt}\langle\Tr {\cal O}\rangle = 0,\ {\cal O} \in {\cal B}_{L-2} \\
&\text{6. Double-trace steady state constraints: } \langle \Tr {\cal O}' \rangle {d\over dt}\langle\Tr {\cal O}\rangle = 0,\ {\cal O}' \in {\cal B}_{L_1},\ {\cal O} \in {\cal B}_{L_2}, \\
&\text{with } L_1 + L_2 \leq L - 2 \\
&\text{7. Restrict to constraints involving only single- and double-traces in the above} \\
&\text{8. Convex relaxation: replace double-traces with } w_{u,v} \text{ and impose } {\cal W}^{(L)} \succeq 0
\fe

The resulting minimum (maximum) obtained from \textbf{SDP}($L$) provides a rigorous lower (upper) bound on the value of $\langle\Tr{\cal Q}\rangle$ that any steady state can realize. We use \texttt{MOSEK} \cite{MOSEK} and \texttt{SDPA-DD} \cite{sdpaddweb,5612693,sdpa,sdpaManual} to solve \textbf{SDP}($L$).\footnote{We use default parameters for \texttt{MOSEK}. For \texttt{SDPA-DD}, we use epsilonStar = 1.0E$-9$, lambdaStar = 1.0E$3$, omegaStar = 2.0, lowerBound = $-1.0$E$5$, upperBound = 1.0E$5$, betaStar = 0.1, betaBar = 0.2, gammaStar = 0.9, and epsilonDash = 1.0E$-9$.} \texttt{MOSEK} is a double-precision solver that is efficient and sufficient for most purposes. However, there are cases where higher precision is required for numerical stability, in which case we use the double-double precision solver \texttt{SDPA-DD}.

In the presence of a global symmetry, there are additional bootstrap constraints that may be imposed. For discrete symmetries, we may impose that the steady state is invariant simply by requiring the expectation values of operators in non-singlet representations to vanish. For continuous symmetries with generators ${\cal G}_\alpha$, we may impose that the steady state is weakly symmetric by
\ie\label{eqn:weakSym}
\langle[{\cal G}_\alpha,{\cal O}]\rangle = 0, \quad {\cal O} \in {\cal B}_L,
\fe
while imposing strong symmetry can be done via
\ie\label{eqn:strongSym}
\langle{\cal G}_\alpha{\cal O}\rangle = 0, \quad {\cal O} \in {\cal B}_L.
\fe
When the strong symmetry constraints (\ref{eqn:strongSym}) are imposed, it may happen that the bootstrap problem \textbf{SDP}($L$) does not admit a feasible point, which can be detected by solvers such as \texttt{MOSEK} and \texttt{SDPA-DD}. In such cases, we conclude that no strongly symmetric steady state exists for the system under consideration, and in particular, that the steady state—if it exists—must be a mixed state.\footnote{To be precise, when SDP solvers produce infeasibility certificates, such as \texttt{pdINF} in \texttt{SDPA-DD}, this should be regarded only as numerical \textit{evidence} for infeasibility, rather than a mathematically rigorous proof. For example, \texttt{SDPA-DD} treats $(x,\textbf{X},\textbf{Y})$ as variables, where $\textbf{Y}$ is the dual variable appearing in the dual problem of maximizing $\tr(F_0\textbf{Y})$ subject to $\tr(F_i\textbf{Y})=c_i$ and $\textbf{Y}\succeq0$, and always keeps the positivity constraints $\textbf{X}\succeq0$ and $\textbf{Y}\succeq0$ satisfied while attempting to minimize the residuals $||\textbf{X}-\sum_{i=1}^mF_ix_i-F_0||$ and $\sum_{i=1}^m(\tr(F_i\textbf{Y})-c_i)^2$, which serve as indicators of infeasibility, while simultaneously reducing the duality gap via Newton-type methods. During this process, however, step lengths may be significantly reduced to maintain positivity constraints even when the residuals or duality gap remain relatively large, and such behavior is then interpreted by the solver as evidence of infeasibility. Another relevant point is that the scale of the initial starting point is set by SDPA parameters, such as lambdaStar, which in general should be taken large enough (see 6.2 of \cite{sdpaManual} for the precise definition of \texttt{pdINF}, which indicates infeasibility only within a subregion of $\textbf{X}\succeq0$ and $\textbf{Y}\succeq0$ determined by the SDPA parameters). To ensure that one does not miss a potentially very small feasible region in this procedure, higher-precision solvers are desirable for infeasibility tests, which is why we used \texttt{SDPA-DD} for such tests.}

\subsection{Nonequilibrium critical points of 1-MQM with an inverted double-well potential}\label{sec:doubleWell}
Consider a quartic potential subject to one-quantum dissipation:
\ie\label{eqn:doubleWell}
{d\over dt}\rho=-i[H,\rho]+\gamma\left(-{1\over2}\{\Tr(BA),\rho\}+\sum_{i,j}A_{ij}\rho B_{ji} \right),
\fe
with
\ie
H=\Tr\left( {1\over2}(P^2+X^2)+{g\over N}X^4 \right).
\fe
When $\gamma = 0$, this model has been studied in great detail. An interesting feature of this Hamiltonian is that, as long as $g \geq g_c = -{\sqrt{2}\over6\pi}$, it admits a normalizable ground state in the planar limit. When $g < 0$, the potential becomes an inverted double-well that is unbounded from below. However, tunneling is suppressed at large $N$, so metastable states supported inside the well between the two maxima of the potential can become stable if $g \geq g_c$. For $g < g_c$, however, the system can no longer accommodate normalizable steady states.

In the presence of $\gamma > 0$, we can ask whether steady states may exist even for $g < g_c$. In fact, the solvable example of the inverted quadratic potential in section \ref{sec:quadraticEx} suggests that strong dissipation leads to the existence of normalizable steady states for unbounded potentials that would otherwise not support them. Therefore, given a value of $g < g_c$, we expect a nonequilibrium critical point $\gamma_c(g)$ such that no normalizable steady state exists for $\gamma < \gamma_c(g)$, while steady states do exist for $\gamma \geq \gamma_c(g)$. We also remark that even at finite $N$, such nonequilibrium critical points should exist for any value of $g < 0$. At fixed $g$, the critical value of $\gamma$ is expected to be a decreasing function of $N$.

We can obtain lower bounds on $\gamma_c(g)$ in the planar limit as follows. Given a set of values $g = g_* \leq g_c$ and $\gamma = \gamma_*$, we solve \textbf{SDP}($L$) in (\ref{SDPDefinition}) using \texttt{SDPA-DD} with any objective function. If \texttt{SDPA-DD} returns \texttt{pdINF}, indicating primal-dual infeasibility, it implies that \textbf{SDP}($L$) at $g = g_*$ and $\gamma = \gamma_*$ is not feasible, and therefore, no normalizable steady state exists. This yields a rigorous lower bound $\gamma_* < \gamma_c(g_*)$ on the nonequilibrium critical point. A similar strategy was recently used in \cite{Cho:2024kxn} to derive lower bounds on the critical temperature at which thermal equilibrium ceases to exist, in the case $g_c \leq g < 0$.

In practice, we also add symmetry constraints to \textbf{SDP}($L$). The first symmetry is the $\mathbb{Z}_2$ symmetry, which acts as $(X, P) \rightarrow (-X, -P)$ and can be implemented simply by setting the expectation values of words of odd length to zero. The second symmetry is the weak $U(N)$ symmetry given in (\ref{eqn:weakSym}). We denote by \textbf{SDPsym}($L$) the SDP problem combining \textbf{SDP}($L$) together with these symmetry constraints. If we find \textbf{SDPsym}($L$) to be infeasible, it follows that \textbf{SDP}($L$) is also infeasible. This is because the existence of any steady state implies the existence of a steady state that respects all the symmetries.

\begin{figure}
  \centering
  \includegraphics[width=0.48\textwidth]{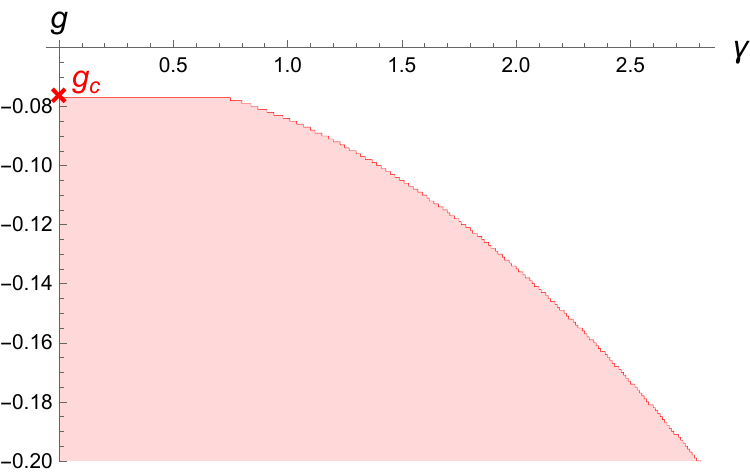}\hfill
  \includegraphics[width=0.48\textwidth]{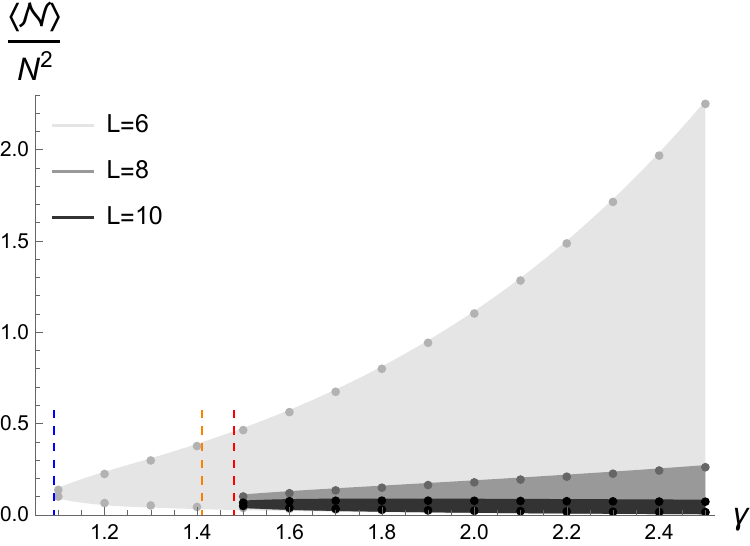}
  \caption{
  \textbf{Left:} Region in the $(\gamma, g)$-plane where \texttt{SDPA-DD} finds \textbf{SDPsym}($L=8$) for the 1-MQM system (\ref{eqn:doubleWell}) to be infeasible is colored in red. No normalizable steady state exists in this region. Also shown is the critical value $g_c$ at $\gamma = 0$. 
  \textbf{Right:} \textbf{SDPsym}($L$) upper and lower bounds on ${\langle{\cal N}\rangle \over N^2}$ for $L=6$ (light gray), $L=8$ (gray), and $L=10$ (black) as functions of $\gamma$ at $g = -0.1$ for the 1-MQM system (\ref{eqn:doubleWell}), obtained using \texttt{SDPA-DD}. Shaded regions are allowed by bootstrap, while the unshaded regions are excluded. The tightest lower bounds on $\gamma_c(g = -0.1)$ from the infeasibility of \textbf{SDPsym}($L$) at $L=6$, $8$, and $10$ are $\gamma = 1.09$, $1.41$, and $1.48$, shown as blue, orange, and red dotted lines, respectively.
  }
  \label{fig:invDW}
\end{figure}

In Figure \ref{fig:invDW}, we present results of \textbf{SDPsym}($L$). In the left panel, \texttt{SDPA-DD} returned \texttt{pdINF} for \textbf{SDPsym}($L=8$) in the red region of the $(\gamma, g)$-plane, implying that no normalizable steady state exists there.\footnote{We applied \texttt{SDPA-DD} to \textbf{SDPsym}($L$) on a finely spaced grid of $(\gamma, g)$ values, where $g$ ranged from $-0.2$ to $-0.077$ in increments of $0.001$. At fixed $g$, we increased $\gamma$ by $0.01$ until the \texttt{SDPA-DD} result was no longer \texttt{pdINF}.} The red boundary line thus provides rigorous lower bounds on $\gamma_c(g)$ at each value of $g$. To assess how close these lower bounds are to the actual values of $\gamma_c(g)$, we compare the tightest lower bounds on $\gamma_c(g = -0.1)$ obtained from \textbf{SDPsym}($L$) for $L = 6$, $8$, and $10$, which were $\gamma = 1.09$, $1.41$, and $1.48$, respectively. These are shown as dotted lines in the right panel, where \textbf{SDPsym}($L$) bounds on ${\langle{\cal N}\rangle \over N^2}$ for $L = 6$, $8$, and $10$ are also displayed. For feasible values of $\gamma$, we observe that the upper and lower bounds converge rapidly as $L$ increases, suggesting a unique steady state that respects all imposed symmetries.\footnote{Even when we did not impose the weak $U(N)$ and $\mathbb{Z}_2$ symmetry constraints, \textbf{SDP}($L=6$) produced the same bounds as \textbf{SDPsym}($L=6$).} When the strong $U(N)$ symmetry constraints (\ref{eqn:strongSym}) are further added to \textbf{SDPsym}($L=6$), \texttt{MOSEK} finds the problem to be infeasible at all tested values of $\gamma$, implying that the unique steady state is not strongly symmetric under $U(N)$ and is therefore a mixed state.

\begin{figure}
  \centering
  \includegraphics[width=0.48\textwidth]{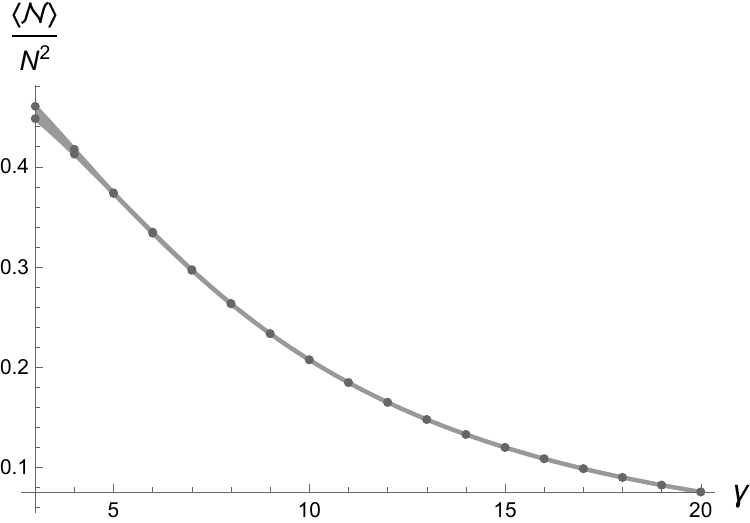}\hfill
  \includegraphics[width=0.48\textwidth]{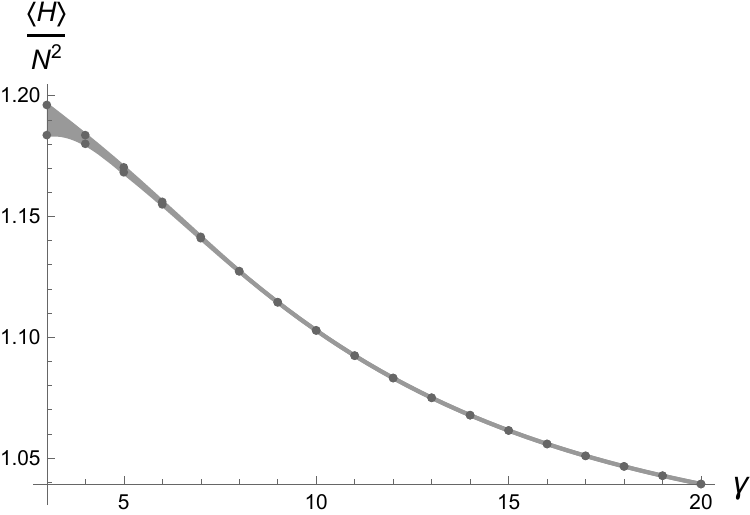}
  \caption{
  \textbf{SDPsym}($L=10$) upper and lower bounds on ${\langle{\cal N}\rangle \over N^2}$ (left) and ${\langle H \rangle \over N^2}$ (right) as functions of $\gamma$ for the 1-MQM system (\ref{eqn:doubleWell}) at $g = 2$, obtained using \texttt{MOSEK}.
  }
  \label{fig:g2DW}
\end{figure}

Before closing this subsection, we also present \textbf{SDPsym}($L=10$) bootstrap bounds on ${\langle{\cal N}\rangle \over N^2}$ and ${\langle H\rangle \over N^2}$ at $g = 2$ in Figure \ref{fig:g2DW}. We observe that both quantities decrease as $\gamma$ increases. This behavior is expected, since at large $\gamma$, the dissipative term—whose unique steady state is the Fock vacuum—dominates over the Hamiltonian term. When we further impose the strong $U(N)$ symmetry constraints (\ref{eqn:strongSym}), \texttt{MOSEK} finds \textbf{SDPsym}($L=6$) to be infeasible at all tested values of $\gamma>0$. Therefore, the steady state is not strongly symmetric under $U(N)$ and is thus a mixed state.

\subsection{Preliminary analysis of 2-MQM with unbounded potential}
\begin{figure}
  \centering
  \includegraphics[width=0.55\textwidth]{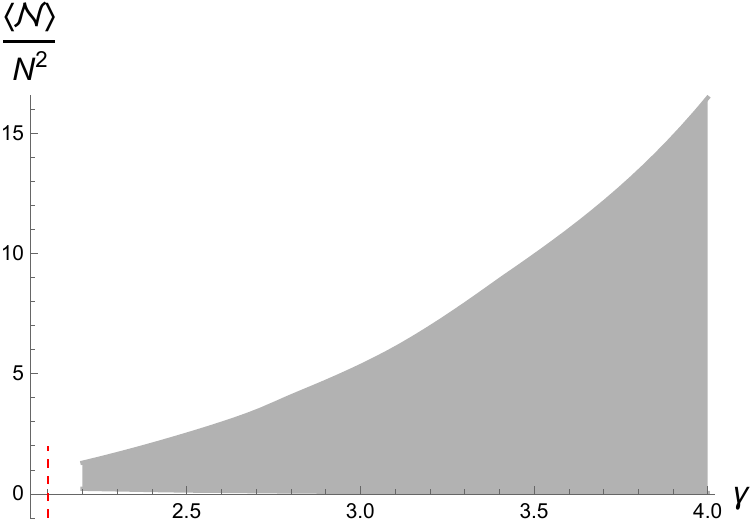}
  \caption{
  \textbf{SDPsym2}$(L=6)$ upper and lower bounds on ${\langle{\cal N}\rangle\over N^2}$ for the 2-MQM example (\ref{eqn:2MQMLind}) at $g=-1$, obtained using \texttt{SDPA-DD}. The red dotted line indicates $\gamma=2.1$, a lower bound on the nonequilibrium critical point $\gamma_c(g=-1)$.
  }
  \label{fig:2MQM}
\end{figure}

The analysis of the previous section can be extended straightforwardly to multi-MQM systems. We proceed to a preliminary analysis of the following 2-MQM Lindblad equation:
\ie\label{eqn:2MQMLind}
{d\over dt}\rho=-i[H,\rho]+\gamma\sum_{I=1,2}\left(-{1\over2}\{\Tr(B_IA_I),\rho\}+\sum_{i,j}A_{I,ij}\rho B_{I,ji} \right),
\fe
where
\ie
A_{I,ij}={1\over\sqrt{2}}(X_{I,ij}+i P_{I,ij}),~~~B_{I,ij}=A_{I,ji}^\dagger={1\over\sqrt{2}}(X_{I,ij}-i P_{I,ij}),
\fe
and
\ie
H = {1\over2}\Tr((P_I)^2+(X_I)^2)-{g\over4N}\Tr[X_I,X_J]^2.
\fe
The $U(N)$ symmetry generator is given by $G_{ij}=i[X_I,P_I]_{ij}+2N\delta_{ij}$, while the $SO(2)$ symmetry generator is given by ${\cal K}=\Tr(X_1P_2-X_2P_1)$. Since ${\cal K}$ is itself a single-trace operator, the strong $SO(2)$ symmetry constraint simply amounts to setting $\langle{\cal K}\rangle=0$ due to large $N$ factorization, while the weak $SO(2)$ symmetry constraints are given by $\langle[{\cal K},{\cal O}]\rangle=0$.

Typically, only $g>0$ cases are considered, since the Hamiltonian is bounded from below. In the presence of dissipation, we can also consider $g<0$ and ask about the nonequilibrium critical point $\gamma_c(g)$ such that there exists a normalizable steady state if and only if $\gamma\geq\gamma_c(g)$. Employing a similar strategy as in the previous section, we consider the 2-MQM version of the bootstrap problem \textbf{SDP}$(L)$, where variables and constraints are restricted to those involving expectation values of words composed of $X_1, X_2, P_1, P_2$ whose lengths are less than or equal to $L$. We further impose that the density matrix is weakly symmetric under all of $U(N)$, $SO(2)$, and $\mathbb Z_2: \{X_I, P_I\} \rightarrow \{-X_I, -P_I\}$, and denote the corresponding bootstrap problem by \textbf{SDPsym2}$(L)$. In Figure \ref{fig:2MQM}, \textbf{SDPsym2}$(L=6)$ bounds on ${\langle{\cal N}\rangle\over N^2} = {\langle\Tr(P_I^2+X_I^2-2)\rangle\over 2N^2}$ at $g=-1$ are presented. In particular, \texttt{SDPA-DD} finds \textbf{SDPsym2}$(L)$ infeasible for $\gamma \leq 2.1$, leading to the lower bound $2.1 < \gamma_c(g=-1)$.

\section{Matrix quantum optics}\label{sec:MQO}
We now turn our attention to 1-MQM Lindblad equations that are motivated by recently discovered nonequilibrium phase transitions in quantum optics for driven-dissipative Kerr resonators.

\subsection{Nonequilibrium phase transition in matrix quantum optics with one-quantum pumping}
We start by briefly reviewing quantum optics examples exhibiting phase transitions in the thermodynamic limit,\footnote{See \cite{PhysRevA.94.033841,PhysRevA.95.012128,PhysRevA.98.042118} for a more complete discussion.} which motivate the 1-MQM examples to be studied shortly. We consider a cavity containing photons subject to two competing effects. The first is dissipation due to spontaneous emission of photons into the environment. The second is a drive mechanism due to external pumping of photons into the cavity.

To be specific, consider a single bosonic particle quantum mechanical system describing the driven-dissipative Kerr resonator:
\ie\label{eqn:KerrLindblad}
{d\over dt}\rho=-i[h,\rho]+\gamma\left(a\rho a^\dagger-{1\over2}\{a^\dagger a,\rho\} \right),
\fe
with the Hamiltonian
\ie\label{eqn:KerrHamiltonian}
h=\Delta a^\dagger a+\chi\sqrt{\bar N}(a^\dagger+a)+{\omega\over \bar N}\left(a^\dagger\right)^2a^2,
\fe
where $\Delta$, $\chi$, and $\omega$ denote the cavity-pump detuning, one-quantum pumping amplitude, and Kerr nonlinearity respectively, and we are working in a reference frame rotating at the pumping frequency. The Kerr nonlinearity represents self-interactions of quanta inside the cavity, while the pumping term linear in $a^\dagger$ and $a$ represents a coherent source of quanta. We also note that $\Delta$ may take either sign, as it is defined as the difference between the cavity and pump frequencies.

$\bar N$ may not seem to have an intrinsic physical meaning, but it provides a notion of a ``thermodynamic'' limit. At any finite $\bar N$, the system admits a unique steady state. As $\bar N$ increases while keeping all other parameters fixed, the expectation value of the number operator $\langle n\rangle=\langle a^\dagger a\rangle$ scales linearly with $\bar N$. For fixed order-one values of $\Delta<0$, $\chi>0$, and $\omega>0$, one finds a first-order phase transition as $\gamma$ is varied in the strict $\bar N\rightarrow\infty$ limit \cite{PhysRevA.94.033841,PhysRevA.95.012128,PhysRevA.98.042118,Beaulieu2025}. There exists a critical value $\gamma_c$ such that the number of steady states is one for $\gamma\geq\gamma_c$, while it becomes infinite for $\gamma<\gamma_c$. The quantity ${\langle n\rangle\over \bar N}$ serves as an order parameter, which exhibits a discontinuous jump across $\gamma=\gamma_c$.

We readily observe that the large $\bar N$ scaling of (\ref{eqn:KerrHamiltonian}) closely resembles the large $N$ scaling of MQM. This motivates us to consider the following 1-MQM Lindblad equation, which we refer to as \emph{matrix quantum optics}:
\ie\label{eqn:MQO}
{d\over dt}\rho=-i[H,\rho]+\gamma\left(-{1\over2}\{\Tr(BA),\rho\}+\sum_{i,j}A_{ij}\rho B_{ji} \right),
\fe
with the Hamiltonian
\ie
H=\Tr\left(\Delta BA+\chi\sqrt{N}(B+A)+{\omega\over N}B^2A^2\right).
\fe
Analogously to the driven-dissipative Kerr resonator, this Lindblad equation describes a cavity containing adjoint quanta subject to both coherent pumping and spontaneous emission. The key question is whether these competing effects give rise to nonequilibrium phase transitions similar to those in the Kerr resonator. When $\gamma$ is large, we expect a unique steady state resembling the Fock vacuum. In contrast, at $\gamma=0$, there are infinitely many steady states arising from ensembles of Hamiltonian eigenstates. We thus ask whether there exists a nonzero value of $\gamma$ at which the number of steady states changes from one to multiple.

\begin{figure}
  \centering
  \includegraphics[width=0.48\textwidth]{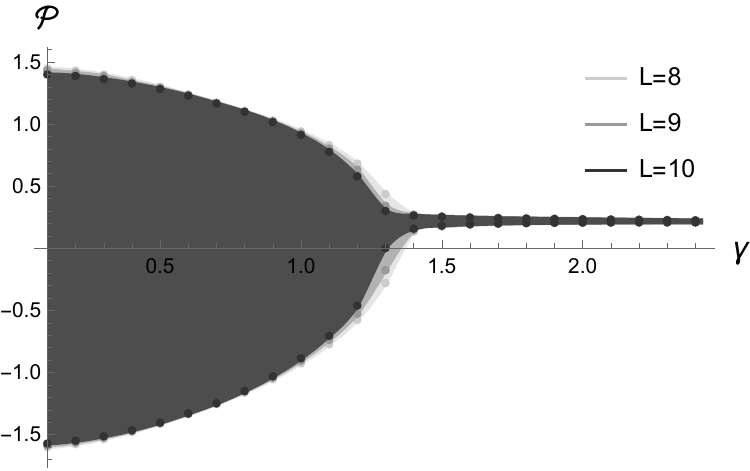}\hfill
  \includegraphics[width=0.48\textwidth]{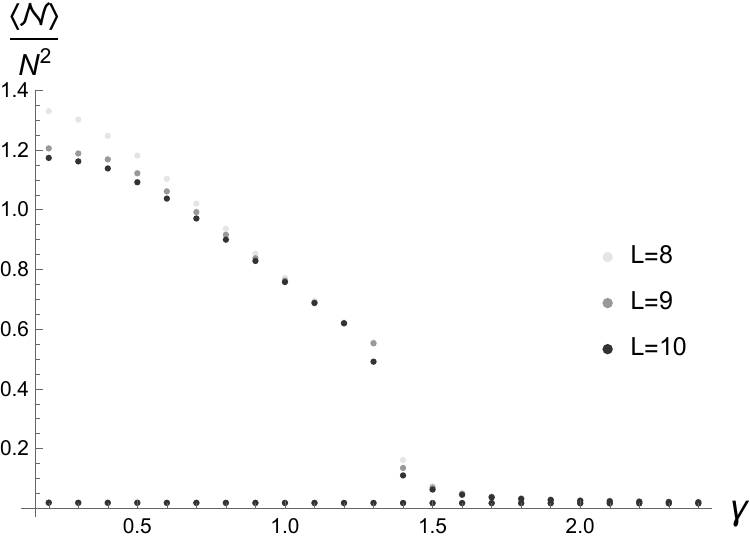}
  \caption{\textbf{SDP}($L$) upper and lower bounds on ${\cal P}$ (left) and ${\langle{\cal N}\rangle\over N^2}$ (right) for $L=8$ (light gray), $L=9$ (gray), and $L=10$ (dark gray), at different values of $\gamma$ with $\Delta=-4$, $\chi={1\over2}$, and $\omega=5$ for the matrix quantum optics system (\ref{eqn:MQO}), obtained using \texttt{MOSEK}.}
  \label{fig:ABPTall}
\end{figure}

We obtain bootstrap bounds on 
\ie\label{eqn:definitionP}
{\cal P}={1\over N^{3/2}}\langle\Tr\left(B+A\right)\rangle~~~\text{and}~~~{\langle{\cal N}\rangle\over N^2}={\langle\Tr{BA}\rangle\over N^2},
\fe
by solving \textbf{SDP}($L$) for the system (\ref{eqn:MQO}) at $\Delta=-4,~\chi={1\over2},~\omega=5$, as functions of $\gamma$. The results are presented in Figure \ref{fig:ABPTall}, where we find compelling evidences for a nonequilibrium phase transition at $\gamma=\gamma_c\approx1.3$. For $\gamma>\gamma_c$, bootstrap upper and lower bounds converge to each other as $L$ increases, implying that there is a unique steady state. In contrast, upper and lower bounds on both ${\cal P}$ and ${\langle{\cal N}\rangle\over N^2}$ start to deviate significantly from each other as we decrease $\gamma$ past the critical value $\gamma_c$, suggesting the existence of multiple steady states for $\gamma<\gamma_c$.

Another question we can address is the existence of steady states which are strongly symmetric under $U(N)$. Firstly, \textbf{SDP}($L$) lower bounds on ${\langle{\cal N}\rangle\over N^2}$ presented in Figure \ref{fig:ABPTall} are strictly greater than zero, implying that they do not correspond to Fock vacuum. Secondly, we can define another bootstrap problem \textbf{SDP2}($L$) by adding strong $U(N)$ symmetry constraints (\ref{eqn:strongSym}) to the bootstrap problem \textbf{SDP}($L$). We find from \texttt{SDPA-DD} that \textbf{SDP2}($L=10$) is infeasible for all values of $\gamma$ that we tested, implying that there is no steady state that is strongly symmetric under $U(N)$. Meanwhile, adding weak $U(N)$ symmetry constraints (\ref{eqn:weakSym}) to \textbf{SDP}($L$) did not change bootstrap bounds at all, suggesting that steady states maximizing or minimizing ${\cal P}$ or ${\langle{\cal N}\rangle\over N^2}$ are mixed states weakly symmetric under $U(N)$.

\begin{figure}[t]
  \centering
  \includegraphics[width=0.48\textwidth]{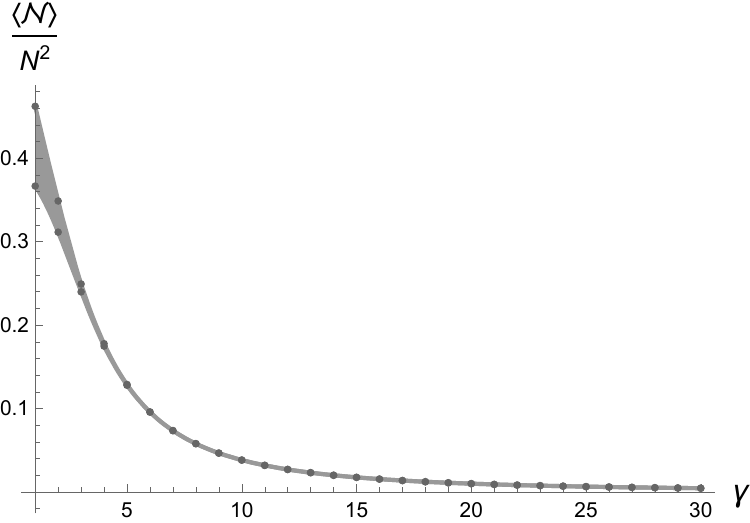}\hfill
  \includegraphics[width=0.48\textwidth]{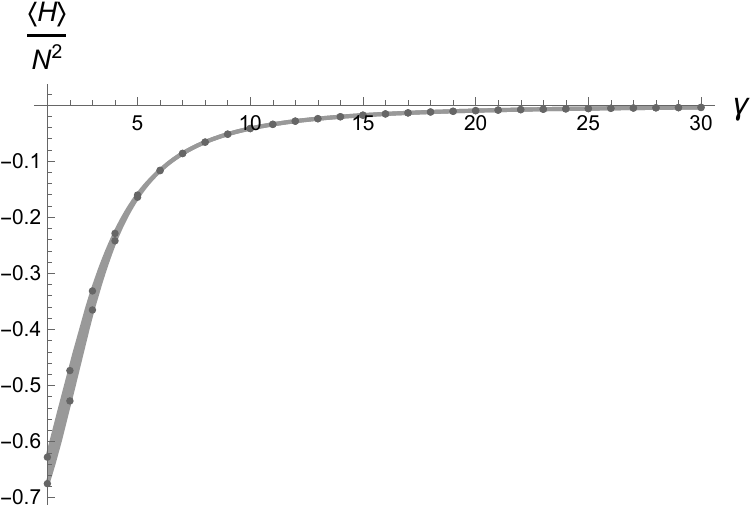}
  \caption{\textbf{SDP}($L=8$) upper and lower bounds on ${\langle{\cal N}\rangle\over N^2}$ (left) and ${\langle H\rangle\over N^2}$ (right) as functions of $\gamma$ at $\Delta=1$, $\chi=1$, and $\omega=1$ for the matrix quantum optics system (\ref{eqn:MQO}), obtained by \texttt{MOSEK}.}
  \label{fig:ABPlot}
\end{figure}

We also consider $\Delta>0$ where no obvious phase transition is observed. In Figure \ref{fig:ABPlot}, we present bootstrap bounds on ${\langle{\cal N}\rangle\over N^2}$ and ${\langle H\rangle\over N^2}$ obtained by solving \textbf{SDP}($L=8$) at different values of $\gamma$, for $\Delta=1,~\chi=1,~\omega=1$. When strong $U(N)$ symmetry constraints (\ref{eqn:strongSym}) are further imposed, \texttt{SDPA-DD} produced \texttt{pdINF}, and we conclude that there is no steady state strongly symmetric under $U(N)$.

\subsection{Nonequilibrium phase transition in matrix quantum optics with two-quantum pumping}
The driven-dissipative Kerr resonator can also exhibit phase transitions in the presence of two-quantum pumping and emission as demonstrated in \cite{PhysRevA.94.033841,PhysRevA.98.042118,Beaulieu2025}. Consider the Lindblad equation
\ie\label{eqn:KerrLindbladZ2}
{d\over dt}\rho=-i[h,\rho]+\gamma\left(a\rho a^\dagger-{1\over2}\{a^\dagger a,\rho\} \right)+{\eta\over\bar N}\left(a^2\rho \left(a^\dagger\right)^2-{1\over2}\{\left(a^\dagger\right)^2 a^2,\rho\}\right),
\fe
with the Hamiltonian
\ie\label{eqn:KerrHamiltonianZ2}
h=\Delta a^\dagger a+{\chi\over2}\left(\left(a^\dagger\right)^2+a^2\right)+{\omega\over \bar N}\left(a^\dagger\right)^2a^2,
\fe
where $\chi$ term corresponds to two-quantum pumping and $\eta$ term corresponds to two-quantum spontaneous emission. There is a $\mathbb Z_2$ symmetry $a\rightarrow-a$. It has been shown that as one varies $\chi$ at fixed values of $\Delta>0,\omega>0,\gamma>0,$ and $\eta>0$, there is a second-order phase transition associated to the $\mathbb Z_2$-symmetry breaking in the large $\bar N$ limit. There exists $\chi_c$ such that Fock vacuum is the unique steady state for $\chi\leq\chi_c$, while nontrivial $\mathbb Z_2$-invariant steady states appear in addition to the Fock vacuum for $\chi>\chi_c$.\footnote{See Appendix \ref{apx:kerrBoot} for a more detailed discussion, where the bootstrap approach presents a sharp nonanalytic behavior at $\chi=\chi_c$.} The order parameter ${\langle n\rangle\over \bar N}$ changes continuously across $\chi_c$. For $\Delta<0$, a first-order phase transition appears as $\chi$ is varied where ${\langle n\rangle\over \bar N}$ changes discontinuously across the critical value of $\chi$.

We define an analogous matrix quantum optics system as
\ie\label{eqn:MQOZ2}
{d\over dt}\rho=-i[H,\rho]+\gamma\left(-{1\over2}\{\Tr(BA),\rho\}+\sum_{i,j}A_{ij}\rho B_{ji} \right)+{\eta\over N}\left(-{1\over2}\{\Tr(B^2A^2),\rho\}+\sum_{i,j}A^2_{ij}\rho B^2_{ji} \right),
\fe
with the Hamiltonian
\ie
H=\Tr\left(\Delta BA+{\chi\over2}(B^2+A^2)+{\omega\over N}B^2A^2\right).
\fe
In terms of the time evolution of expectation values, this system is equivalently described by
\ie
{d\over dt}\langle {\cal O}\rangle &= ~i\langle[H,{\cal O}]\rangle
\\
&+\gamma\left({1\over2}\langle[\Tr(BA),{\cal O}]\rangle+\sum_{i,j}\langle B_{ij}[{\cal O},A_{ji}]\rangle\right)+{\eta\over N}\left({1\over2}\langle[\Tr(B^2A^2),{\cal O}]\rangle+\sum_{i,j}\langle B^2_{ij}[{\cal O},A^2_{ji}]\rangle\right).
\fe
The system admits a $\mathbb{Z}_2$ symmetry, under which $A_{ij} \rightarrow -A_{ij}$. If we are specifically interested in $\mathbb{Z}_2$-invariant steady states, we may impose that the expectation values of all words of odd length vanish. We denote the corresponding semidefinite program by \textbf{SDP}${}_{\mathbb{Z}_2}(L)$.

\begin{figure}
  \centering
  \includegraphics[width=0.55\textwidth]{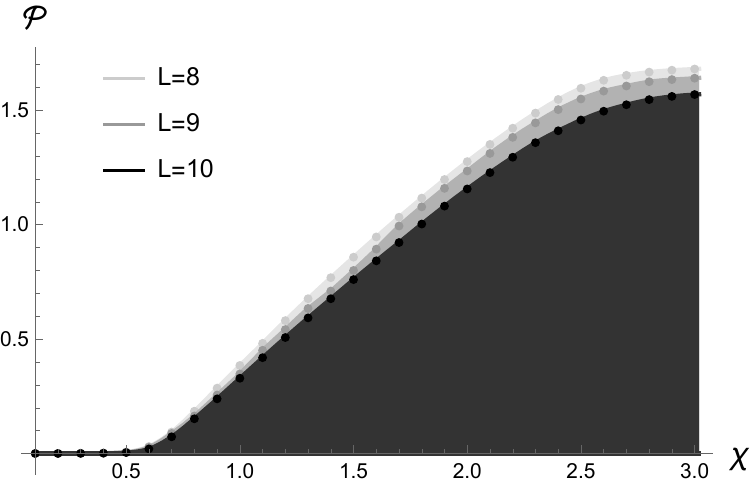}
  \caption{\textbf{SDP}($L$) upper bounds on ${\cal P}$ obtained by \texttt{MOSEK} at $L=8$ (light gray), $L=9$ (gray), and $L=10$ (black), for the matrix quantum optics system (\ref{eqn:MQOZ2}). Parameter values are $\Delta=-10$, $\omega=5$, $\gamma=1$, and $\eta=1$.}
  \label{fig:ABZ2PT}
\end{figure}

We start with $\Delta<0$. In Figure \ref{fig:ABZ2PT}, we present bootstrap upper bounds on ${\cal P}$ from \textbf{SDP}($L$) as functions of $\chi$, the two-quantum pumping amplitude, at $\Delta=-10$, $\omega=5$, $\gamma=1$, and $\eta=1$. (Lower bounds are simply the negatives of the upper bounds due to the $\mathbb{Z}_2$ symmetry.) Note in particular that we did not impose $\mathbb{Z}_2$ symmetry constraints, which would have automatically set ${\cal P}=0$.

We find compelling evidence of a $\mathbb{Z}_2$-breaking nonequilibrium phase transition at $\chi = \chi_c \approx 0.6$. For $\chi < \chi_c$, ${\cal P}$ remains close to zero, suggesting that steady states are $\mathbb{Z}_2$-symmetric. In contrast, for $\chi > \chi_c$, the upper bounds begin to deviate significantly from zero as $\chi$ increases, indicating the emergence of steady states that are not symmetric under $\mathbb{Z}_2$. Although the bounds have not fully converged as $L$ increases from $L=8$ to $L=10$, a nearly nonanalytic behavior around $\chi = \chi_c$ appears at all $L$ values shown in the figure. We expect that the upper bounds will eventually converge to nonzero values of ${\cal P}$ for $\chi > \chi_c$, realized by steady states that break the $\mathbb{Z}_2$ symmetry.

\begin{figure}
  \centering
  \includegraphics[width=0.55\textwidth]{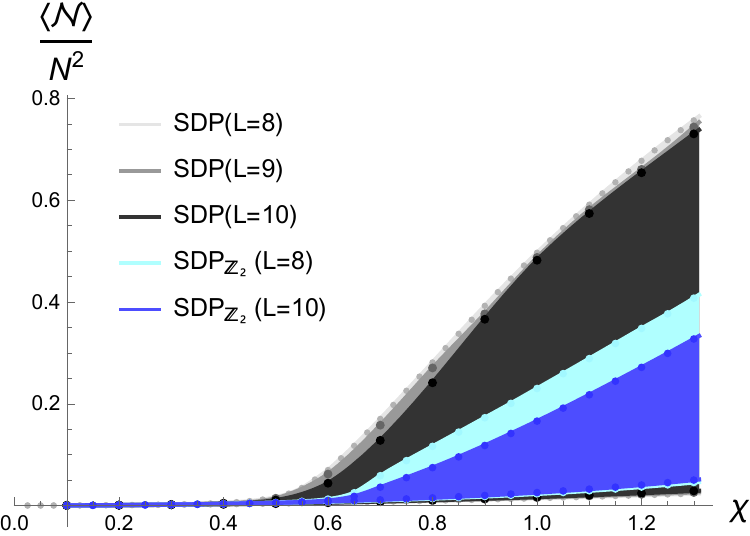}
  \caption{Bootstrap bounds on ${\langle{\cal N}\rangle\over N^2}$ for the matrix quantum optics system (\ref{eqn:MQOZ2}) with $\Delta=-10$, $\omega=5$, $\gamma=1$, and $\eta=1$, obtained by \texttt{MOSEK}. Different colors correspond to: \textbf{SDP}($L=8$) (light gray), \textbf{SDP}($L=9$) (gray), \textbf{SDP}($L=10$) (black), \textbf{SDP}${}_{\mathbb{Z}_2}(L=8)$ (light blue), and \textbf{SDP}${}_{\mathbb{Z}_2}(L=10)$ (blue).}
  \label{fig:ABZ2PTN}
\end{figure}

We also present bootstrap bounds on ${\langle{\cal N}\rangle\over N^2}$, both with and without $\mathbb{Z}_2$ symmetry constraints, in Figure \ref{fig:ABZ2PTN}. We observe that even after imposing $\mathbb{Z}_2$ symmetry constraints, the bootstrap bounds exhibit a qualitative change around the critical point $\chi \sim \chi_c$. Moreover, the apparent differences between the upper bounds obtained from \textbf{SDP}($L$) and \textbf{SDP}${}_{\mathbb{Z}_2}(L)$ for $\chi > \chi_c$ provide further evidence for the existence of steady states that are not invariant under $\mathbb{Z}_2$. Note that if the upper and lower bounds from \textbf{SDP}${}_{\mathbb{Z}_2}(L)$ do not converge as $L$ increases, it suggests the possibility of multiple steady states that are $\mathbb{Z}_2$-invariant, reminiscent of the Kerr resonator case (\ref{eqn:KerrLindbladZ2}). However, higher-$L$ results appear necessary to draw a definitive conclusion. Finally, we remark that the lower bounds on ${\langle{\cal N}\rangle\over N^2}$ remain strictly positive across the parameter range studied, indicating that the Fock vacuum is not a steady state. This stands in contrast to the Kerr resonator case (\ref{eqn:KerrLindbladZ2}), where the Fock vacuum always persists as a steady state in the large $\bar{N}$ limit.

\begin{figure}
  \centering
  \includegraphics[width=0.48\textwidth]{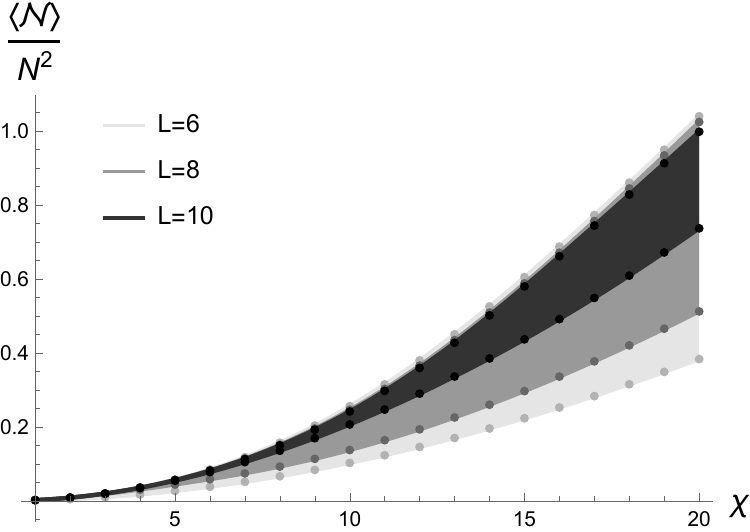}\hfill
  \includegraphics[width=0.48\textwidth]{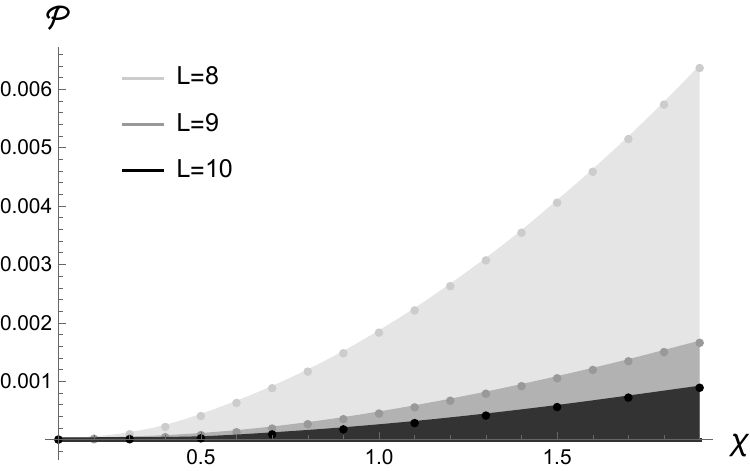}
  \caption{
    \textbf{Left:} \textbf{SDP}${}_{\mathbb{Z}_2}(L)$ bounds on ${\langle{\cal N}\rangle\over N^2}$ for $L=6$ (light gray), $L=8$ (gray), and $L=10$ (black) as functions of $\chi$. 
    \textbf{Right:} \textbf{SDP}($L$) upper bounds on ${\cal P}$ for $L=8$ (light gray), $L=9$ (gray), and $L=10$ (black) as functions of $\chi$. 
    Both plots were obtained using \texttt{MOSEK} at $\Delta=10$, $\omega=5$, $\gamma=1$, and $\eta=1$ for the matrix quantum optics system (\ref{eqn:MQOZ2}).
  }
  \label{fig:ABZ2PosDPlot}
\end{figure}

For the case of $\Delta>0$, we find no evidence of a phase transition. Both \textbf{SDP}${}_{\mathbb Z_2}(L)$ bounds on ${\langle{\cal N}\rangle\over N^2}$ and \textbf{SDP}($L$) bounds on $\cal P$, presented in Figure \ref{fig:ABZ2PosDPlot}, show no indication of nonanalytic behavior. The convergence of \textbf{SDP}${}_{\mathbb Z_2}(L)$ bounds on ${\langle{\cal N}\rangle\over N^2}$ as $L$ increases suggests the existence of a unique $\mathbb Z_2$-invariant steady state. Furthermore, the \textbf{SDP}($L$) upper bounds on $\cal P$ approach zero with increasing $L$, indicating the absence of any steady state that breaks $\mathbb Z_2$ symmetry.

Finally, we note that for all cases considered in this subsection, adding strong $U(N)$ symmetry constraints (\ref{eqn:strongSym}) rendered the SDP problems infeasible, whereas adding weak $U(N)$ symmetry constraints (\ref{eqn:weakSym}) did not affect the bootstrap bounds. We therefore expect that the steady states realizing the bootstrap bounds in this subsection are mixed states that are weakly symmetric under $U(N)$.

\section{Discussion}\label{sec:discussion}
Nonequilibrium physics, in general, has been much less explored compared to its equilibrium counterpart. However, recent developments across various areas of physics consistently suggest that nonequilibrium settings are not only natural but also rich in novel physical phenomena. In this context, nonequilibrium MQM offers an exciting opportunity to investigate nonequilibrium aspects of quantum gravity. The results presented in this work indicate the presence of genuine nonequilibrium phase transitions in MQM, which merit further investigation.

\textbullet~ It would be desirable to obtain bootstrap results incorporating additional constraints, not only by increasing the cutoff $L$, but also by including nonlinear constraints involving multi-trace operators. The bounds presented in this work on nonequilibrium phase transitions in matrix quantum optics have not yet fully converged. It is therefore important to incorporate further constraints to numerically confirm convergence and place the existence of these nonequilibrium phase transitions on a firmer footing. One possible approach is to scan over the values of a few single-trace expectation values, which can significantly reduce the number of independent multi-trace variables, as demonstrated in \cite{Lin:2024vvg,Lin:2025srf}.

There remains a logical possibility that the inclusion of additional constraints could, in principle, lead to the conclusion that a unique steady state exists for all nonzero values of the dissipation strength in matrix quantum optics examples. Nonetheless, the nonanalytic behaviors of the bootstrap bounds presented in this work strongly suggest the existence of nontrivial nonequilibrium phase transitions. A related question is whether the full nonlinear bootstrap formulation, incorporating all ingredients of large $N$ MQM, can be shown to converge to an existing steady state. While convergence theorems are well-established for various convex optimization problems,\footnote{See, e.g., \cite{Cho:2023ulr,Fawzi:2023fpg} for convergence proofs in the context of classical and quantum many-body systems on infinite lattices.} such convergence has not been investigated in detail for the present case of large $N$ MQM, where the problem is inherently nonconvex.

\textbullet~ Although more involved, the bootstrap approach can be straightforwardly extended to gauged MQM, where alternative methods based on suitable truncation schemes may also be available. A relevant remark is that, in the regime where a semiclassical Einstein gravity picture emerges, there is expected to be no essential difference between the gauged and ungauged MQM \cite{Maldacena:2018vsr}. Away from such a regime, the two versions of MQM generally differ. The results presented in this work suggest that such differences may also be present in the presence of Lindblad dissipative dynamics.

\textbullet~ We have not derived the Lindblad form of MQM from any microscopic Hamiltonian involving both the system and the environment. It would be fascinating to obtain such a derivation and identify instances where dual gravity descriptions are available. At the same time, regardless of whether such a microscopic derivation is straightforward or not, it is worth emphasizing that the setups discussed in this work are interesting in their own right and may offer nontrivial insights into the study of quantum gravity—analogous to how coupling a theory of gravity to a heat sink has proven instrumental in the investigation of quantum extremal surfaces \cite{Penington:2019npb,Almheiri:2019psf}. 

\textbullet~ We implemented the bootstrap method to study nonequilibrium MQM systems in this work. While this approach does not prove the existence of specific steady states, it can rule out candidate states from the space of valid steady states. It would be valuable to develop alternative methods for investigating nonequilibrium MQM, such as the quantum trajectory method \cite{Daley04032014}, which could provide real-time evolution data. A major challenge in applying such methods to MQM lies in the rapid growth of computational costs as the matrix size increases.

\textbullet~ We also note that, even within the bootstrap framework, there are several straightforward directions for further study in nonequilibrium MQM. When an absorbing state exists—which is a pure steady state—it is possible to use the bootstrap method to obtain rigorous bounds on either the gap of the Liouvillian superoperator (when the absorbing state is the unique steady state), or on certain ratios involving differences of expectation values between an extremal steady state and the absorbing state \cite{Cho:2025dgc} (when nontrivial steady states exist in addition to the absorbing state). However, we have yet to find an example of nonequilibrium MQM that exhibits such an absorbing phase transition. The driven-dissipative Kerr resonator with two-quantum pumping and emission in the thermodynamic limit provides an example of an absorbing phase transition, as the Fock vacuum is always a steady state. As we have seen in this work, however, its matrix version does not possess any pure steady state. Lastly, irrespective of the existence of an absorbing state or any steady state, it is straightforward to bootstrap the real-time evolution of expectation values in nonequilibrium MQM, following the ideas presented in \cite{Lawrence:2024mnj}. It would be interesting to study how a generic initial state evolves into a nontrivial steady state.

\section*{Acknowledgments}
We would like to thank Aashish Clerk, Barak Gabai, Antal Jevicki, Yue-Zhou Li, Henry Lin, Andrew Pocklington, Shinsei Ryu, Xi Yin, and Zechuan Zheng for helpful discussions, and Antal Jevicki for comments on the draft. We are especially grateful to the authors of \cite{shinsei} for sharing their draft and kindly agreeing to coordinate our submissions. This work is supported by Clay C\'ordova's Sloan Research Fellowship from the Sloan Foundation.

\appendix
\section{Bootstrap approach to the thermodynamic limit of the Kerr resonator}\label{apx:kerrBoot}
In this appendix, we briefly discuss the bootstrap approach to the strict thermodynamic limit of the driven-dissipative Kerr resonator, focusing on the example (\ref{eqn:KerrLindbladZ2}). We begin by introducing the rescaled creation and annihilation operators:
\ie
\tilde a= {a\over\sqrt{\bar N}},~~\tilde a^\dagger= {a^\dagger\over\sqrt{\bar N}},
\fe
so that
\ie
[\tilde a,\tilde a^\dagger]={1\over\bar N}.
\fe
To avoid cluttered notation, we drop the tilde $\sim$ on $\tilde a$ and $\tilde a^\dagger$ from this point onward, and let $a$ and $a^\dagger$ denote the rescaled operators. Then, the Lindblad equation (\ref{eqn:KerrLindbladZ2}) becomes
\ie\label{eqn:resKerrLindbladZ2}
{d\over dt}\rho=\bar N\left(-i[{\tilde h},\rho]+\gamma\left(a\rho a^\dagger-{1\over2}\{a^\dagger a,\rho\} \right)+{\eta}\left(a^2\rho \left(a^\dagger\right)^2-{1\over2}\{\left(a^\dagger\right)^2 a^2,\rho\}\right)\right),
\fe
with the rescaled Hamiltonian
\ie\label{eqn:resKerrHamiltonianZ2}
{\tilde h}=\Delta a^\dagger a+{\chi\over2}\left(\left(a^\dagger\right)^2+a^2\right)+{\omega}\left(a^\dagger\right)^2a^2.
\fe

Expectation values of strings composed of $a$ and $a^\dagger$ are now of order ${\bar N}^0$, whereas a single commutator between them is suppressed by ${\bar N}^{-1}$. The time evolution of operator expectation values is given by:
\ie\label{eqn:resKerrOp}
{d\over dt}\langle{\cal O}\rangle={\bar N}\left(i\langle [{\tilde h},{\cal O}]\rangle+\gamma\left({1\over2}\langle[a^\dagger a,{\cal O}]\rangle+\langle a^\dagger[{\cal O},a]\rangle \right)+\eta\left({1\over2}\langle[\left(a^\dagger a\right)^2,{\cal O}]\rangle+\langle \left(a^\dagger)\right)^2[{\cal O},a^2]\rangle\right) \right).
\fe
For example, when ${\cal O}=a^\dagger a$, we obtain
\ie
{d\over dt}\langle{a^\dagger a}\rangle=-\gamma\langle a^\dagger a\rangle-2\eta\langle \left(a^\dagger \right)^2a^2\rangle-i\chi\left(\langle \left(a^\dagger\right)^2\rangle-\langle a^2\rangle \right).
\fe
In the strict $\bar N \to \infty$ limit, we retain only the ${\bar N}^0$ terms on the right-hand side (RHS) of (\ref{eqn:resKerrOp}). This explains why the Fock vacuum is always a steady state of (\ref{eqn:resKerrLindbladZ2}). To see this explicitly, we evaluate the RHS of (\ref{eqn:resKerrOp}) in the Fock vacuum. The only surviving terms on the RHS are
\ie
i {\bar N} {\chi\over2}\langle[\left(a^\dagger\right)^2+a^2,{\cal O}]\rangle.
\fe
This expression is nonzero only if ${\cal O}$ is either $\left(a^\dagger\right)^2$ or $a^2$. Consider the former case:
\ie
i {\bar N} {\chi\over2}\langle[a^2,\left(a^\dagger\right)^2]\rangle=i {\chi}\langle a a^\dagger\rangle={i\chi\over\bar N},
\fe
which is of order $\bar N^{-1}$. Thus, we confirm that in the strict ${\bar N}\rightarrow\infty$ limit, the RHS of (\ref{eqn:resKerrOp}) vanishes when evaluated in the Fock vacuum, as expected.

We can similarly understand why the Fock vacuum ceases to be a steady state when the Hamiltonian includes the one-quantum pumping term $a^\dagger+a$. In that case, an analogous contribution to the RHS of (\ref{eqn:resKerrOp}) evaluated in the Fock vacuum becomes
\ie
\sim i {\bar N} \langle[a,a^\dagger]\rangle=i,
\fe
which is of order $\bar N^0$ and therefore survives in the $\bar N\rightarrow\infty$ limit.

\begin{figure}
  \centering
  \includegraphics[width=0.7\textwidth]{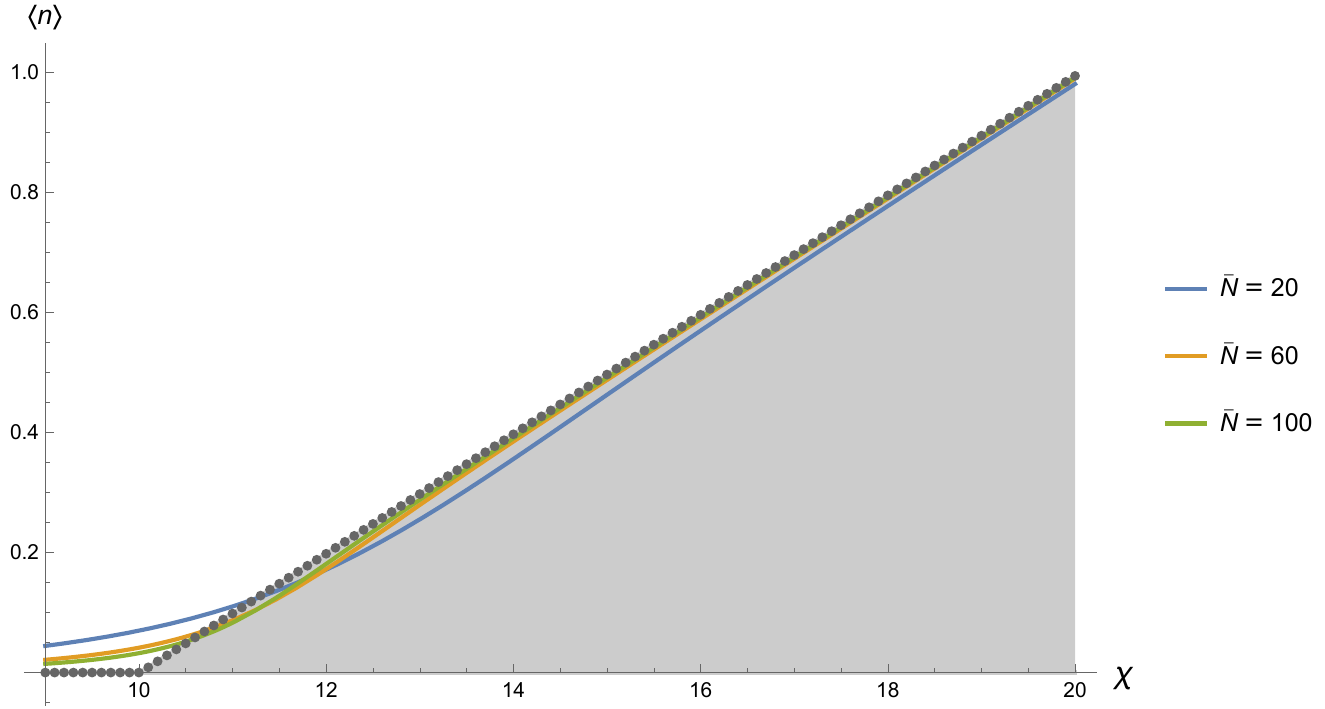}
  \caption{\texttt{MOSEK} bootstrap bounds (gray) on $\langle n\rangle=\langle a^\dagger a\rangle$ for steady states of (\ref{eqn:resKerrOp}) in the strict thermodynamic limit $\bar N\rightarrow\infty$, evaluated at different values of $\chi$. The bounds were obtained from constraints involving words of lengths up to 6, with parameters $\Delta=10,~\omega=5,~\gamma=1,~\eta=1$ and $\mathbb Z_2$ symmetry assumption. The gray-shaded region is allowed by the bootstrap. Also shown are exact solutions \cite{PhysRevA.94.033841} at ${\bar N}=20, 60,$ and $100$, plotted in blue, orange, and green respectively.}
  \label{fig:kerBoot}
\end{figure}

Steady state constraints ${d\over dt}\langle{\cal O}\rangle=0$ for (\ref{eqn:resKerrOp}) at order ${\bar N}^0$, combined with the positivity condition $\langle{\cal O}^\dagger{\cal O}\rangle\geq0$, define the bootstrap problem for steady states in the strict thermodynamic limit, which can be solved via SDP. We further impose $\mathbb{Z}_2$ symmetry by setting the expectation values of strings of $a$ and $a^\dagger$ with odd length identically to zero. In Figure \ref{fig:kerBoot}, we present bootstrap bounds on $\langle n\rangle=\langle a^\dagger a\rangle$ obtained from constraints involving words of length up to 6, for $\Delta=10,~\omega=5,~\gamma=1,~\eta=1$. In fact, the bounds remained unchanged even when the maximum word length was increased to 8, indicating that the bounds have converged to the exact steady-state values. We also display the exact solutions at finite $\bar N$ from \cite{PhysRevA.94.033841}. These results demonstrate the power of the bootstrap method in capturing the phase transition of the quantum optical system in the strict thermodynamic limit, as demonstrated by the sharp nonanalytic behavior of the upper bound at the critical point $\chi=\chi_c=10$.

\bibliographystyle{JHEP}
\bibliography{noneq}

\end{document}